\documentclass{article}
\usepackage[utf8]{inputenc}
\usepackage{url}
\usepackage{url}
\usepackage{authblk}

\usepackage{graphicx}
\usepackage[export]{adjustbox}
\usepackage{amsmath}
\usepackage[margin=1.5in]{geometry}
\usepackage{amsfonts}

\usepackage{longtable}
\usepackage[T1]{fontenc}
\usepackage{makecell}

\title{Black-box model risk in finance}
\author[1]{Samuel N. Cohen \thanks{samuel.cohen@maths.ox.ac.uk}}
\author[2]{Derek Snow \thanks{dsnow@turing.ac.uk}}
\author[3]{Lukasz Szpruch \thanks{l.szpruch@ed.ac.uk}}
\affil[1]{Mathematical Institute, University of Oxford}
\affil[1,2,3]{The Alan Turing Institute}
\affil[3]{School of Mathematics, University of Edinburgh}

\date{\today \thanks{We are grateful to Katia Babbar for helpful comments and suggestions on a draft of  this paper. We acknowledge the support of the Alan Turing Institute under EPSRC grant no. EP/N510129/1}}

\usepackage{hyperref}
\usepackage[english]{babel}

\usepackage{natbib}
\usepackage{chapterbib}

\usepackage[inline]{enumitem} 

\usepackage{amsthm}

\usepackage[english]{babel}

\usepackage{hyperref}
\usepackage[utf8]{inputenc}
\usepackage[dvipsnames]{xcolor} 

\usepackage{etoolbox}

\usepackage{makeidx}

\begin{document}

\maketitle
\begin{abstract}
    Machine learning models are increasingly used in a wide variety of financial settings. The difficulty of understanding the inner workings of these systems, combined with their wide applicability, has the potential to lead to significant new risks for users; these risks need to be understood and quantified. In this sub-chapter, we will focus on a well studied application of machine learning techniques, to pricing and hedging of financial options. Our aim will be to highlight the various sources of risk that the introduction of machine learning emphasises or de-emphasises, and the possible risk mitigation and management strategies that are available.
\end{abstract}


\section{Introduction}
     Traditionally, the tractability of pricing and hedging methods was arguably more critical than their accuracy, and the limits of computation determined what methods were useful. The Black--Scholes formula is concise, simple to understand, can be implemented on a handheld calculator \citep{lo2019adaptive}; these features were critical to its wide adoption. Similarly, the Heston model benefits from convenient (fast) Fourier transformation methods (see, for example, \citep{gatheral2006}) and the SABR model from a convenient approximation (see \citep{hagan2002managing, obloj2007fine}), which have formed a key part of their attractiveness. While many more sophisticated and accurate models have been developed, computational bottlenecks impeded their wider adoption.

In recent years, machine learning models in finance have become streamlined; in just a few lines of packaged code, modellers can develop state-of-the-art models with online computing power and open-source software \citep{snow2020machine,dixon2020machine}.
However,  the risks of blindly using machine learning solutions, without understanding their inner workings and inherent drawbacks, are significant.  

In this sub-chapter, we seek to give an overview of the key issues which arise when using machine learning in finance, and some remedies which have been suggested. Rather than focus on developing a particular algorithm, we take a higher-level view of the risks and challenges which arise in these contexts. We wish to highlight that machine learning is not a panacea for financial markets, instead it provides tools which allow practitioners to shift between different sources of risk, some of which have not been a primary concern in the past. 

We will focus on those risks which are a core part of machine learning -- the risks inherent in data and in the modelling algorithms used. We will not discuss what the WEF calls the erosion of ``human financial talent'' where humans lose the skill to challenge and disagree with machine learning systems \citep{mcwaters2019navigating}, although this is potentially a significant concern in many financial applications.

There are two broad uses of machine learning in finance. The first application of machine learning in finance is to remove computational barriers and enable use of advanced models in day to day business operations. When used in this manner, `machine learning' is providing a next generation computational tools, which are used to speed up and improve traditional modelling. For example, when calibrating an option pricing model one often needs to price many derivatives many times, using a variety of potential parameter values -- this is a task that can be improved by using a machine-learnt approximation for the pricing operator.  \citet{hutchinson1994nonparametric} trained a neural network on simulated data to learn the Black--Scholes option pricing formula. A number of efficient algorithms have recently been developed to approximate parametric pricing operators with flexible modelling assumptions (for example, see \citet{horvath2020deep,jacquier2019deep,ferguson2018deeply,mcghee2018artificial,vidales2018unbiased,sabate2020solving}). This in turn can eliminate the calibration bottlenecks commonly found in using realistic pricing models.

The second application involves a more fundamental change in the approach to modelling and working with data, where traditional, low-dimensional, handcrafted models are replaced with abstract over-parameterized models, using the tools of machine learning to avoid overfitting. These methods may be used to represent the statistical features of underlying assets, to determine prices, hedges and risk properties of portfolios in terms of market observables, as well as a combination of these tasks. This application depends in a far more significant way on the historical data available, leading to various challenges: for example, it becomes hard to understand what is driving the price of a derivative, and the data modelling and preprocessing steps might introduce an additional set of risks, which can be a cause of unease for regulators and risk managers.

In this subchapter, for the sake of concreteness, we focus our attention on the challenge of pricing and hedging derivatives, which we outline in Section \ref{sec:pricingproblem}, and principally on the use of one machine learning method (deep neural networks) in this challenge. In Section \ref{sec:data} we discuss issues connected with the sources of data that are used as inputs into machine learning algorithms, while in Section \ref{sec:models} we are concerned with the risk associated with the way probabilistic modelling is incorporated within machine learning.

\section{A practical application of machine learning} \label{sec:pricingproblem}

To get more acquainted with a neural network solution, we will take a closer look at the problem of pricing and hedging an exotic derivative, by trading in a financial market. Here we give a non-technical description; for a technical primer on neural networks see, for example, the work of \citet{bengio2017deep}. The inputs to our problem will be a combination of historical market data and commonly accepted handcrafted models (depending on the precise approach), the latter may be used to generate additional simulated data for training purposes. The key outputs are prices, hedging strategies and risk assessments for exotic options and portfolios of exotic options and other assets. 

\subsection{How to use Neural nets for derivative modelling}
\begin{figure}[h!]
     \includegraphics[width=\textwidth,left]{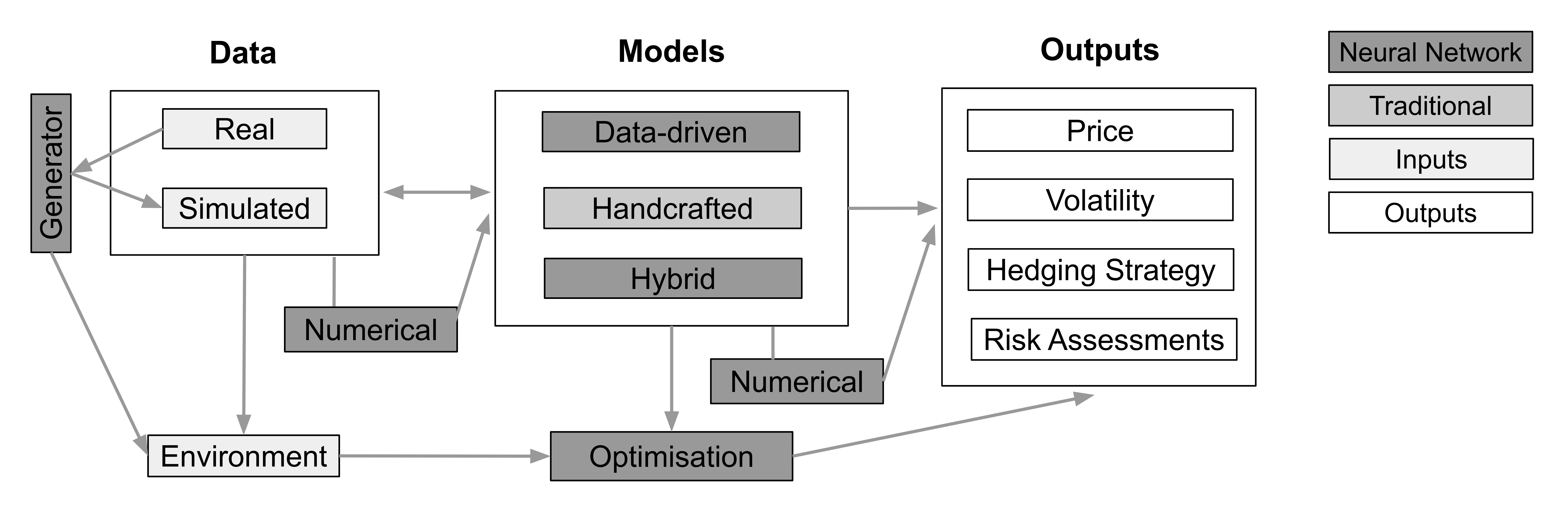}
       \caption{\emph{Neural Network Pricing and Hedging}. This figure illustrates five broad roles for neural networks as part of developing pricing, volatility, and hedging models. Neural networks can learn directly from real data (Data-driven), can be used as a numerical or computational tool (Numerical), can enhance handcrafted models (Hybrid), can generate novel simulated data (Generator), and can be used in reinforcement learning models to develop strategies in a dynamic way (Optimisation).}
    \label{fig:neural}
\end{figure}

The precise role of machine learning in options pricing and the data used to support it can vary significantly. If we consider one particular class of machine learning methods -- neural networks -- in Figure \ref{fig:neural} we present one way of classifying some applications of this method, looking at whether they principally are concerned with the processing and generation of data, or with building models for financial markets, and how these contribute to different outputs.

 Many of the use-cases for neural networks boil down to their ability to learn complex, high dimensional, non-linear relationships; for example, to solve partial differential equations in high dimension, to develop data-driven models with large feature sets, and to find optimal policies in large state-spaces via reinforcement learning. The effectiveness of neural nets in high dimensional settings suggests they have ability to overcome the computational curse of dimensionality\footnote{It is widely conjectured that neural nets are effective in situations where the problem at hand admits an accurate low dimensional representation, however, this representation is not known a priori \citep{fefferman2016testing} }.  We divide the neural network use-cases into five broad modes of application: Data-driven models, Hybrid models, Numerical approximations, Online Optimisation methods, and Generator models.

 \begin{enumerate}[label={(\arabic*)}]

 \item \textbf{Numerical approximations} are based on traditional parametric models, and exploit neural networks to, for example,  approximate pricing or hedging functionals in the form of solutions to parametric families of PDEs. This approximation can significantly speed up calibration problems. In these applications, the neural network is not trained against real-world historical data, but is used purely to approximate complex functions, in an efficient way. The key difficulty in this area is the calibration of hyperparameters and network architectures, and the implementation to industrial standards.

 \begin{itemize}

 \item Some models require computationally expensive procedures involving solving a partial differential equation (PDE) or performing Monte Carlo simulations to estimate the option price, implied volatility, or hedging strategy. For these models we can use offline neural networks to efficiently approximate an entire pricing or hedging function \citep{hutchinson1994nonparametric}. 
 
 \item	Some applications depend on solving a high-dimensional and/or non-linear PDE, even if underlying model is simple. For example to price and hedge path-dependent options or to compute XVAs\footnote{This is similar to the now classic use of regression in the Longstaff--Schwarz approach to pricing American options \citep{longstaff2001}, where neural networks can give a more flexible approximating class of functions.}. Neural networks can be used as a function approximation tool which works well in high dimension, and are particularly efficient at solving PDEs when blended with Monte Carlo simulation. \citep{barucci1997neural,beck2019deep,sabate2020solving}
 
\item  Some problems, in particular in calibration of handcrafted models (for example, Heston or SABR models), require repeated calculation of various option prices prices under a variety of parameters. By providing an efficient means of approximating this calculation for a range of parameter choices, neural networks speed up the process of calibration, allowing a more efficient use of data.  \citep{andreou2010generalized,bayer2019deep,vidales2018unbiased}. These methods often depend on simulating option values from the handcrafted model, under a range of parameter values.
 
 \end{itemize}

\item \textbf{Data-driven models} rely on real market data to approximate pricing and hedging functions. These models disregard handcrafted models in their entirety and simply use historical, synthetic or simulated data of any type to learn new relationships and features \citep{ghaziri2000neural,montesdeoca2016extending}.

\item \textbf{Hybrid models} rely on historical, simulated, or synthetic data to approximate pricing and hedging functions and also constrain or impose knowledge onto the architecture of an otherwise unconstrained neural network.

 \begin{itemize}
\item Some models first leverage a handcrafted model to estimate prices and then build a data-driven model to learn the difference or residuals between the observed price and the handcrafted model estimate \citep{lajbcygier1997improved}.

\item Other models constrain a universal neural network by adding domain knowledge to the architecture to learn more realistic relationships that increases the interpretability or efficiency of the model e.g., forcing monotonous relationships towards one direction by adding penalties to the loss function \citep{garcia2000pricing,dugas2009incorporating,gierjatowicz2020robust}.
\end{itemize}

\item \textbf{Online Optimization methods}: A number of option types, for example American options, benefit from learning optimal stopping rules using neural networks in a reinforcement learning framework; others may benefit from learning a value function or a hedging strategy that benefits from temporal optimal control, e.g. a model that takes evolving market frictions into account in an environment or control system \citep{buehler2019deep, kolm2019dynamic}. 

\item \textbf{Generator models} can take any data as input and generate new data that has the same statistical properties. Data can be generated by applying a calibrated `handcrafted' model (or from a range of handcrafted models) or from a machine learning generative model. Alternatively, they may be learn from data observed in one situation to generate representative data in a related setting. The first of these uses (where the generated data matches statistical properties of historical observations) is called `synthetic data', and is a subset of `simulated data', which includes scenarios that were not present in the historical data. The generated data’s purpose is principally to aid the performance of machine learning pipelines, for example to provide an environment to train further models with reinforcement learning. It's worth noting that the generator, and hence the simulated data, should be seen as a statistical model\footnote{Here the network architecture, loss function and training method of the generator are all modeling choices.} for our observations. This approach can be viewed as model-based data boosting  \citep{buhler2020data,ni2020conditional, marinari2019}. 

\end{enumerate}

Using a combination of these approaches, we can now build an abstract pipeline for learning to price and hedge options.

\begin{enumerate}[label={(\arabic*)}]
\item Using historical data and current market data, build up a collection of training trajectories of the assets under consideration, as well as a representation of the state of the market. As we typically only have one trajectory of past data, one often needs to augment historical observations with models or simulated data. We have discussed two main approaches to this:

\begin{enumerate}[label={(\roman*)}]
\item Design and train \emph{generative} models to provide additional realistic data, or provide a rich parametric class with which to work.
\item Train \emph{handcrafted} models (possibly using neural networks as a \emph{numerical} tool to speed up calibration) from which to simulate.
\end{enumerate}
\item Using this data, either
\begin{enumerate}[label={(\roman*)}]
\item Use reinforcement learning to \emph{optimize} hedging strategies (and thus determine initial prices), using the simulator as a training environment
\item Learn \emph{data-driven} pricing relationships and hedging strategies by observing prices in historical data.
\item Learn a \emph{hybrid} model that first trains on simulated data, and then transfers this learning over to real data for efficient training. 
\item Use a further \emph{numerical} approach, to solve the PDEs arising from the (possibly high dimensional) models. Equivalently, one can ensure the calibrated model generates trajectories with probabilities from a risk-neutral measure, and use Monte Carlo simulation to estimate prices.
\end{enumerate}
\end{enumerate}

\subsection{Black box trade-offs}
The pipeline we have outlined above gives a very flexible approach to modelling and pricing of financial derivatives, however this is not a `free lunch'. Neural networks are notoriously opaque as a modelling tool, and are often implemented simply as a `black-box' approach to function approximation (the hybrid models discussed above being a partial exception to this). An important practical question is whether the potential disadvantages of black-box hedging can be justified by increased performance, and whether the risks associated with this approach can be distinguished and quantified.

An industry standard to assess the quality of a new model is to compare it with simpler benchmarks such as BS-delta hedging with the presence of transnational costs  \citep{davis1993european,whalley1999optimal}.  There is preliminary evidence that suggests, at least in simple, constant volatility settings, these benchmark models have performance close to that of reinforcement learning agents \citep{mikkila2020optimal}. If that is true, these benchmarks should be preferred because they have easy to explain analytical solutions.

\citet{ruf2020hedging} have shown, on an out-of-sample test set, that a simple fixed hedging strategy that hedges calls by $ 0.9\ \times\ \delta_{\mathrm{BS}}$ and puts by $ 1.1\ \times\ \delta_{\mathrm{BS}} $, where $\delta_{\mathrm{BS}}$ is the delta under the classic Black--Scholes model, outperformed 14 out of 16 models, including all the supervised neural network models with 1-day rebalancing, and outperformed all models with 2-day rebalancing. It should be noted, that these tests were performed in a simple one period setting, with  no transaction costs. These results do not directly extend to a large basket of derivatives; as a result, more tests are needed. However, these results suggest that more complex black-box models may fail to outperform simpler ones.

It is also possible that improved feature selection and simple models might be a better solution than a direct application of neural networks. \citet{ruf2020hedging} compared a neural network model with a linear regression model  to estimate the hedge ratio, on simulated and real data. Predictors in the linear model included standard model sensitivities under the Black--Scholes model: moneyness, Delta, Vega, implied volatility, time to maturity and Vanna.  They conclude that the classical option sensitivities already contain the non-linearities necessary to build an effective hedging strategy for common options, in a financially significant and efficient way. They further showed that this linear regression model outperformed their neural network model. 

Their approach diverges from \citet{buehler2019deep}, who do not use option sensitivities as variables, but instead rely on the belief that the Greeks indirectly present themselves as non-linear functions that the agent has access to via the market state, in the form of hedging instrument prices. Future experiments should test hedging performance on a basket of derivatives, in a multi-period setting, with dynamic volatility, transaction costs, and environmental feedback. Such a real-life setup could benefit the deep hedging approach \citep{buehler2019deep}, as this has the capacity for direct feedback from the environment and online training.

\section{The role of data}\label{sec:data}
\subsection{Data risks}
We now focus our attention on how the use of machine learning methods highlights risks associated with the underlying financial data. One can identify three primary sources of risk: biases in the training data, erroneous data or erroneous preprocessing, and legal and regulatory data risks. We will not focus on legal and regulatory risks. 
As we are moving the dial from handcrafted towards data-driven models, the data risk increases significantly. On the one hand, handcrafted models are more robust to biases and errors in the data, and the risk of using inadequate models is easier to detect. On the other other hand, for data-driven models the training data becomes an integral part of the model, making them more sensitive to data risk.

\subsubsection{Biased data}
A key issue in financial data is that the majority of data is backward-looking, and there is no guarantee that future behaviour of financial markets will be represented by historical observations. We typically only have one trajectory of historical data -- we cannot see what \emph{might} have happened in different scenarios -- which makes it difficult to build a clear view of the range of likely outcomes in the future. Any recent changes in the true underlying state of a system that are not incorporated in a model’s training dataset will also lead to biased predictions.

These risks are not particular to machine learning -- they are well known issues in financial markets; however, the use of machine learning models, which often depend on observed data in a more significant way than traditional handcrafted models, means understanding and managing these risks is critical for the success of these approaches. Here we summarize some key forms of bias that historical data could exhibit:

\begin{enumerate}[label={(\alph*)}]

    \item \textbf{Backward looking}: most data reflect prices and signals obtained in the past. This means that data could reflect a state of affairs that no longer applies, e.g. an options model might only have access to data from a low volatility period, or from an old regulatory regime. Financial markets are reactive and don't follow universal laws -- for example, the increase of high-frequency trading has changed the nature of many financial markets (see, for example, the discussion in \citet {mackenzie2018}). Restricting to only the very recent past, and projecting a model's predictions only into the near future, can mitigate this concern somewhat, but often results in significantly less data being available for training. 
    
  \item \textbf{Spurious correlations}: for some financial domains it is prudent to record and collect attributes that have some theoretical basis. For example, it is questionable whether an option pricing model should contain sentiment features. It is often difficult to identify intuitively unreasonable relationships within a black-box model, and the increase in dimensionality of the models being used results in a vast increase in the range of potential relationships that could be inferred \citep{fan2016guarding}. Spurious correlations are well known in finance, but the increased use of machine learning techniques can exacerbate this problem.
  \item \textbf{Sample disparity}: biases in the sampling procedure might lead to data that doesn't fairly represent the state of the market. For example, a firm may wish to use the same algorithm for trading over multiple exchanges and geographic locations, each of which has subtly different conventions and data. This introduces biases within the data which can be magnified through the use of machine learning models, particularly when a model is trained in one setting then deployed for use elsewhere. 
  \item \textbf{Imbalanced inputs}: some evidence shows that even when your sample accurately reflects the true state of the market, it remains imbalanced -- rare events may be significant, but are only infrequently represented in historical samples. Many data-driven models are known to favour the performance of the majority outcome, to limit overall model errors \citep{provost2000machine}. Within a financial setting, this might correspond to a model which performs well in low-volatility regimes, even when high-volatility periods are observable (infrequently) within the training data. A related idea is the use of stressed periods for calculating risks within the Basel accords -- these infrequent periods are significant to overall performance, and need to be explicitly taken into account.
  \item \textbf{Insufficient data}: one often has insufficient data to use machine-learning models well. The calibration of neural networks requires significant quantities of data, which are often not available for training in a financial context \citep{gu2018empirical}. The richer the context in which an algorithm is to be run, and the more finely tuned its behaviour needs to be, the larger the quantity of data needed. It is worth emphasising that this is not to say that the data in finance is `small', but that often it is not `large' in the directions needed -- we may have enormous datasets due to high-frequency observations of a large number of asset's order books, but these will be of little use in determining good models over long time periods. 
\end{enumerate}

\subsubsection{Data errors and preprocessing}

A further concern in many applications is that data may display subtle errors, which need to be addressed before it can be effectively used. This is a common concern in many applications of machine learning, and data-cleaning methodologies form a key part of the implementation of these methods.

\begin{enumerate}[label={(\alph*)}]
\item Observed financial data can fail to satisfy fundamental economic constraints, which can be subtle. For example, as discussed in \citet{cohen2020detecting}, historic options price data, for both listed and OTC contracts, may be inconsistent with no-arbitrage constraints, particularly in emerging markets. If such data is na{\"\i}vely used when training a trading system, it is plausible that the system would learn to exploit this apparent arbitrage opportunity. Given these errors could arise due to multiple sources (for example, stale quotes being listed as live in historical data), this can lead to significant error in the resulting learnt behaviour.
\item When working with time-series data, it is critical to respect information flow when e.g splitting data into training, validation, and test sets; engineering features; or normalising data. Errors in this process can `leak' information from the future, leading to unrealistic performance.
\item Financial data often has a particular concern around precise timekeeping, which may not be reflected in the accuracy of the data given. Particularly when working with very high-frequency data, failing to take into account latency and other implementation issues can have a significant effect, which may not be well reflected or available in historic data (see, for example, the effect of latency in \citet{cartea2018shadow}). This is particularly the case with the increased attention being given to non-market data sources (for example signals from online news sources), where historic time-stamping may be of low quality.
\item Financial data is often heavy tailed and not stationary, making it difficult to detect and exclude erroneous data. Typical methods (such as Winsorizing) have the potential to introduce significant bias, particularly when considering extreme events.
\end{enumerate}


\subsection{Data solutions}

There are many process improvements that can be implemented to decrease data risks, for example, performing data quality monitoring, documenting and reviewing the manipulation of input data, and educating and training individuals involved in data manipulation tasks. Another key approach, to fix biased and limited data, is to generate synthetic or simulated data which is free from (or even corrects for) these issues. We will outline two key approaches -- the top-down approach of synthetic data generation, and the bottom-up approach of agent based modelling.

\subsubsection{Synthetic data generation}

Synthetic data generation (SDG) is a top-down data generation solution. It can help to address some of the data biases and errors listed above. It does so by augmenting the quantity and quality of historical data, but it does not attempt to provide a simulator which can model feedback effects for an agent's interventions in a market.

At a high level, a synthetic data generator attempts to build a probabilistic model which would generate observations similar to historical data. Generative models such as generative adversarial networks (GANs) and variational autoencoders (VAEs) have demonstrated great success in seemingly high dimensional setups \citep{wan2017variational, lin2020using}. If used correctly, SDGs could allow for a more comprehensive approach to future-proofing and validating machine learning pipelines; ameliorating some structural deficiencies in data and amending distributional biases \citep{louizos2015variational}.

In a financial context, \citet{takahashi2019modeling} have shown that (GANs) can be used to generate synthetic data that matches most known stylized features of returns; \citet{ni2020conditional}  have shown how mathematically principled feature extraction methods such as signature models can be used to efficiently implement conditional GANs for generic time series data. Related ideas, but combined with VAEs, are presented in \citet{buhler2020data}. \citet{henry2019generative} developed efficient algorithms building upon optimal transport theory and highlighted an interesting application of data generators for detecting anomalies. Algorithms based on restricted Boltzmann machines have been developed in \citet{kondratyev2019market}, who coined the term 'market generators'. An alternative approach is to learn the underlying dynamics of the system, allowing a path to evolve through time -- this is the approach taken by neural-SDE models \citep{gierjatowicz2020robust}. \citet{fu2019time} have shown how conditional GANs can be used to produce synthetic data for different market scenarios. \citet{koshiyama2020generative} have shown how these methods can be used to validate trading strategies.

SDGs still pose a form of modelling risk, the generators are only as good as the data from which it constructs its generating function; building an SDG involves choosing a metric, a loss function, and a training algorithm for parameter selection. As such, SDGs introduce model risks within the data used to train downstream models, and these risks may be difficult to identify depending on the use-case. 

At the present time, research in this area lacks standardised benchmarks and theoretical guarantees. Most off-the-shelf methods are not built with financial applications in mind, and are therefore likely to generate simulated data which exhibits arbitrage or other economically unrealistic phenomena.  Moreover, many of these models remain black-box and are not easily interpretable. 

The key benefit, however, is that these methods are expressive and work in high dimensions. This is the main difference when comparing with traditional methods using handcrafted features. Synthetic data can also be used to generate data according to expert opinions and known facts, e.g., can be conditioned to form the observed volatility smiles. And SDGs generally offer a more accurate and robust oversampling method than traditional methods like SMOTE (Synthetic Minority Oversampling Technique) that simply repeat existing records \citep{chawla2002smote}. They also provide a convenient solution for missing data imputation and outlier treatment \citep{xu2018synthesizing}.

Synthetic data generation tools can be used as part of a larger solution to address some of the most common upstream data errors. They can be used side-by-side with federated learning techniques to improve the quality of single standing resources, by pooling data across, departments, subsidiaries, companies, or data-providers \citep{goetz2020federated}. 

Deep generative models for synthetic data generation remains a new field, and although they have potential to alleviate some of the known issues of neural network models, it is clear that they have the potential to introduce further risks. Overall, as with other methods, they can be seen as shifting
 risks away from the quantity and quality of data, by including probabilistic modelling (with its associated risks) at a very early stage in the analysis pipeline.

\subsubsection{Market simulator engine}

Agent-based model (ABM) simulators, unlike SDGs, are a bottom up data solution and date back to the 1990s. Notable early models include those by \citet{levy1994microscopic} and the Santa Fe Artificial Stock Market \citeyear{palmer1994artificial,arthur1996asset}. ABMs model markets as evolving systems of competing, autonomous interacting agents \citep{lebaron2000agent}. 

The development of ABMs has seen multiple waves of interest. The first wave of market simulators in the 1990s was a deliberate move away from classical economic theories to advance financial market knowledge, the second wave was a reaction to the failure of economic models in foreseeing the financial crises of 2008, the third wave was a call to understand high frequency trading and the flash-crashes in 2010 and 2013, and the fourth and current wave combines the concerns with the past, but emphasises the use of simulators to train machine learning agents.

A key advantage of ABMs is that, as bottom-up models, they attempt to learn the feedback effects of agents acting within the market. This has the advantage that these effects can be modelled, but makes training much more difficult - usually involving explicit modelling decisions, and requiring more data to train. Again we see that the issue of historical data not containing counterfactual histories, or being too limited for our purposes, are being addressed, but doing so introduces increases our reliance on statistical models, rather than on observed data.

 Modeling feedback is important for training environments to be realistic. For example when hedging or trading strategies are trained and tested on historical data, the success of the model still cannot be reliably demonstrated, even when using holdout sets for validation. Training environments with appropriately modelled feedback can, at least partially, mitigate this issue. Such training environments are also critical for deploying on-line reinforcement learning solutions as they allow pre-training of these systems before implementation in the real market. This is critical in applications where the costs and risks of exploration are significant. 

Agent-based modelling has, in recent years, allowed for the design of high-fidelity simulated markets \citep{belcak2020fast, byrd2019abides}. These artificial markets can run millions of in-silico trials to test counterfactual theories, research emergent phenomena, and train and test algorithms.

 A current trend is that quantitative funds are looking to establish risk management systems that develop scenarios with no historical precedent\footnote{In 2017, Jane Street published a technical presentation of their own exchange, motivated to train and test new algorithms and models. It has been reported to handle messages in the rate of 500k/second with latencies in the single digit microseconds \citep{BibEntry2021Jan}.}. With a simulator, one can perform training and backtesting for trading, execution, and placement algorithms under various conditions. Causal assessments can be performed for market impact and market slippage. Lastly, simulators can also be used as a means of generating synthetic data, given that financial data of sufficient granularity is often highly proprietary and/or expensive to access.

\subsubsection{Standardized cleaning and preprocessing methodologies}

Issues surrounding data quality often are specific to the particular use-case. The increasing use of varied data sources, often with little standardization, will inevitably result in the preprocessing of financial data becoming more important. 

Some approaches, for example the no-arbitrage constraints for option books in \citet{cohen2020detecting}, rely on preprocessing data to conform with prescribed characteristics. In this case, given the no-arbitrage constraints restrict the range of possible option prices significantly, imposing these requirements has the potential to address errors coming from a variety of sources. These methods can also be run on data coming from a SDG or ABM, in order to ensure that the simulated data is economically reasonable.

An approach which can serve to highlight potential concerns, is to look at the sampling frequency and periods of data. By comparing the results of using different but comparable datasets, it is possible to gauge the stability of calibration and models, and hence to identify causes for concern. 

More generally, learning from other areas of machine learning, the development of common examples, codebases and resources, in an open-source manner, has the potential to improve the identification and processing of data errors. A significant risk is that inappropriate methods for dealing with data errors will be separately developed, implemented and used, without sufficient oversight or criticism. The use of well-developed, understood, and standardized tools is a key part of modern machine learning, and the development of preprocessing tools appropriate to finance should be seen in this light. 

As part of this, the development of publicly discussed use-cases, with realistic data, would allow for new methods to be evaluated in a consistent manner, and for best-practice to be developed. While this is the case in other areas of machine learning, there is still much scope for improvement when it comes to financial data and problems.

\section{The role of models}\label{sec:models}

\subsection{Model risks}

As we have discussed, the use of machine learning changes, but does not eliminate, the use of classical mathematical modelling. Classical models may appear explicitly in machine learning methods (for example, in a hybrid pricing model), or may be subtly incorporated in the simulations used to support more explicitly data-driven approaches. Typically, however, machine learning methods aim to construct models from flexible (`non-parametric') families, combining the classical tasks of model selection and calibration into a single step. In this section, we will discuss the risks which arise from these modelling decisions, in a machine learning context.

It is worth noting that the presence of model risk depends strongly on how machine learning methods are used. Using machine learning tools for numerical procedures typically introduces little additional model risk, as one can often verify the solutions using other techniques. For example, when using a neural network to estimate option prices, for the sake of quickly calibrating the parameters of a classical model, it is straightforward to verify (using traditional PDE or Monte Carlo methods) that the calibrated model gives the correct prices of those options -- the neural network is only serving as a numerical tool.

Conversely, end-to-end deep reinforcement learning, for example of a  hedging strategy, exposes users to risks in multiple forms: models with too many parameters risk overfitting to available data, leading to both poor performance and a misunderstanding of a model's inaccuracy; the common use of synthetic and simulated data hides an additional layer of model risk in the training environment; complex models are more exposed to reward hacking, poisoning attacks, and other adversarial concerns and are typically less interpretable than simpler models. 

The problem of calculating a price for a financial derivative which is consistent with the market can be seen as equivalent to finding a map that takes market data (e.g. prices of underlying assets, interest rates, prices of liquid options) and returns the no-arbitrage price of the derivative. One way to do this is to select a martingale model (to prevent arbitrage) that can be calibrated to market data, by which we mean that the model matches the observed prices of liquid assets.  

While this is a dominating approach in the industry, the introduction of a model necessarily introduces model risk, and there are infinitely many models that can fit market data.  In the robust finance paradigm, see \citep{hobson1998robust,cox2011robust}, one takes a conservative approach and instead of computing a single price one constructs pricing intervals that are consistent with market data. Without imposing further constrains, the class of all calibrated models might be too large, and consequently, the corresponding pricing intervals too wide to be of practical use \citep{eckstein2019robust}.  It is therefore natural to consider a smaller search space of models (e.g. SDEs with continuous coefficients) and use data and machine learning to select an appropriate model (i.e. the coefficients of the SDE).

This approach has been recently applied in \citep{gierjatowicz2020robust}. The key idea is to use SDEs to describe the model dynamics but, instead of fixing its coefficients, to allow the drift and diffusion to be given by an overparametrized neural network. These `neural SDE' models provide a systematic framework for model selection, but can also produce robust estimates on the derivative prices.

A concern for model risk is not new in finance, but the use of machine learning methods can be seen as typically emphasising some risks over others. In Table \ref{typicalriskstable}, we present an overview of the typical distinctions between handcrafted and machine learning perspectives on model risk. 

\begin{table}[!hp]
\caption{Comparing Typical Risks Between Handcrafted and Machine Learning Methods}\label{typicalriskstable}
\begin{longtable}{|p{25mm}|p{45mm}|p{45mm}|}
\hline \thead{Risk} & \thead{Handcrafted} & \thead{Machine Learning} \\
\hline 
Structural Risk 

& Lower dimensional models which are easy to calibrate, but fail to capture all aspects of the market's behaviour. Generally a higher bias than variance and more prone to underfitting.

& High dimensional models which require large amounts of data to calibrate, but can capture fine detail when fitted well. Can often incorporate new sources of information in a convenient manner. Generally a higher variance than bias and more prone to overfitting.
 \\
\hline 
Model Sensitivity

& Few parameters and model inputs. Model outputs vary smoothly with calibration and input. Well understood sensitivities to erroneous inputs.

& High-dimensional parameters and data inputs. Model outputs can vary sharply with inputs. Sensitivities to erroneous inputs can vary significantly.
\\
\hline 
Adversarial Attacks

& Reasonably robust calibration and not susceptible to data poisoning attacks. Calibration can be easily monitored by users. Adversarial defenses not a key part of most models.
& Susceptible to attacks, require robust training and adversarial defences, but these can be incorporated as a key part of the model. Not easily monitored by users.
\\
\hline 
Model Drift
& Models naturally incorporate economic intuition and underpinnings. Few parameters to update online, but do not often incorporate updating as a core part of the model.
& Model based on data patterns which may change over time. Many parameters need to be updated dynamically, which can lead to unstable behaviour. Model updating can be included as a core part of the approach. \\

\hline
\end{longtable}
\end{table}
\subsubsection{Structural risk}

Within a machine-learning paradigm, one usually combines the stages of model selection and calibration. Given data on a supposed relationship or phenomenon, one aims to directly fit a model to this data with which to predict, simulate and build understanding.

For our example of pricing and hedging of options, we can focus on the task of pricing an option given historical market data. Our data consist of historical observations of market data, and we aim to build a function which can take new observations and provide us with prices in the future. To do this, some basic modelling assumptions are unavoidable:
\begin{itemize}
    \item Does the price of an option depend only on the current market state, on the recent past, or on a long history of market observations. Equivalently, what are the inputs to the pricing function that I wish to find?
    \item Do I wish to make conditional predictions (say of an option price given a stock price) or do I wish to give simulations of both simultaneously?
    \item Does the relationship between market observations and prices remain stable through time? If not, how do I choose training periods which are representative of the situations where I will apply my function in the future?
    \item If the observed prices are not perfectly predicted by market data, so I have noisy observations, are the noises independent, or are they correlated between times and assets?
\end{itemize}
In each case, the answer given to these questions will be incorporated in our machine learning model, and introduces model risk at a structural level.

These general concerns are common to both classical and machine learning methods, however the increased flexibility of machine learning methods may suggest that (as one can include more observations in a model), that they would be less present in a machine learning approach. 

Even after these general concerns are addressed, machine learning methods introduce risks similar to the `model risk' of classical mathematical finance. Within the paradigm of machine learning, models are not chosen explicitly but implicitly, through the choice of training data, training algorithm and the often \emph{ad hoc} choice of a large parametric model (e.g. a neural network and its architecture). Unlike handcrafted methods that are explicitly specified, or hybrid approaches relying on feature engineering, neural networks construct an internal representation of features to capture and approximate functions. 

With neural networks, model specification is not in the direct control of the modeller. Due to this flexibility in feature specification, a larger space of plausible models are explored than in traditional or many other machine learning approaches. The cost of this flexibility is that the model selected may not be the `best' available. Since the fitting of traditional models typically involve solving some convex optimisation problem, a best model can be identified due to the existence of a unique minimum. Neural networks fitting techniques are typically non-convex and many good solutions can be found. 

Adding fuel to the fire, neural networks are known to be sensitive to initialisation conditions \citep{mcmormack1993neural}. Moreover, many sources of randomness are often injected into the training phase of neural networks, this includes the use of dropout (where some neurons are randomly set to zero for network regularisation), early stopping (where the process of gradient descent stops when the performance on a validation set stops improving), and stochastic gradient descent (where random selections of observations are used to fit the network). These additional factors introduce uncertainty in the output of neural network models.
 The injection of noise during training is critical to the performance of these methods, and it leads to, so called, implicit regularisation \citep{neyshabur2017implicit}. That means that stochastic gradient descent methods select regularised solutions, even though regularisation is not explicitly incorporated at the training stage \citep{heiss2019implicit}. In this sense, the model selection step of classical approaches is replaced by the choice of training algorithm, which has a less easily understood connection with model performance. 

Drawing from interpretability research by \citet{lipton2018mythos}, any model's transparency can be broken down in \emph{simulatibility}, \emph{decomposability}, and \emph{algorithmic transparency}. With simulatibility, a human should be able to step through each of the operations in a reasonable time; with decomposability, each part of the model has an intuitive explanation that is understood in isolation;  with algorithmic transparency, there are theoretical guarantees about the behaviour of the algorithm, for example certainty of convergence. Going down this checklist it is clear that neural networks lack simulatibility and decomposability because the parameters in the hidden layers do not have an intuitive explanation. Moreover, for non-convex problems stochastic gradient descent is not guaranteed  to converge. Instead, one can show that the weights of neural networks are represented by Monte Carlo samples from optimal distribution over the parameter space. This perspective allows one to establish convergence guarantees, but does not help with the issue of interpretability \citep{hu2019mean,jabir2019mean}.






\subsubsection{Model sensitivity}

A key selling point of neural networks is their ability to work with high dimensional inputs. However, this comes with a well documented issue of sensitivity, where the learnt relationships vary wildly with small perturbations to the underlying inputs.

Models are known to be fragile when using high dimensional inputs. The reasons are numerous: given the randomness involved in training neural networks, some inputs may spuriously be considered important. This is a particular issue when only limited data is available, or simulated data (from a low dimensional model) is used as training data -- simulated data will typically not explore a full range of market conditions (as it is constrained by the model from which it's generated),  and so the neural net will not learn to provide good answers when novel conditions are encountered. Secondly, when many inputs are used within a model, there is an increased probability that some variables might not be available when a model is put into practice. 

Since the model specification of neural networks is implicit, the modeller and end-user of these methods will often no longer understand how the model has been fit, significantly increasing model specification risk.  Consequently, it is not clear how we can quantify sensitivity of the model. The field is therefore largely left with developing more interpretable model alternatives \citep{nakagawa2019deep} or using post-hoc explanations to assess and visualise what models have learned \citep{li2020beyond}. This however also comes with risk as many post-hoc explanation are not robust and may lead to false sense of security \citep{anders2020fairwashing}. 


\subsubsection{Robustness and adversarial attacks}

The competitive nature of financial markets often leads to particular concerns for machine learning models. As models are used in increasingly automated ways, they need to be able to respond to the pressures placed on them by competitive forces, who have strong incentives to identify and exploit potential weaknesses of a model.

For example, we could consider our challenge of managing an options portfolio, but in a context where market price impact reduces the efficiency of trading. A classic model for order execution with market impact,  \citet{almgren2001optimal}, yields deterministic policies for executing a large buy or sell order, which may have the undesirable effect of `information leakage' (revealing your strategy to other market participants) when used in an illiquid market. In the more complex situation of managing a portfolio, one could consider building a neural network model to perform this task (for optimal execution, a model of this type is given in work by \citet{ning2018double}). The additional randomness of the neural network model would arguably assist in preventing information leakage, when compared to the traditional model. Nevertheless, it is \emph{a priori} unclear whether this additional randomisation would be sufficient, or whether further precautions against information leakage would be needed.

Adversarial attacks can be grouped into many categories, for example, attacks can either be intentional or unintentional. \citet{behzadan2018faults} splits them into attacks on model confidentiality, integrity (does the model behave as intended?), and availability (can the model be disabled by an external actor?). Attacks could also be split into the components that are susceptible to the attack, for reinforcement learning this includes the environment, the observation channel, the reward channel, the decision making system, and the online training system. 

We can consider various way in which a financial reinforcement learning agent can be attacked, with a simple description and illustrative example. 
 These classifications have been adapted from the adversarial threat a matrix developed by MITRE in collaboration with Microsoft, IBM, NVIDIA, and Bosch \citep{kumar2020adversarial}. 

In Table \ref{attackstable}, we present examples of adversarial attacks against a trading system. We first list those which are internal to the company, many of which can arise inadvertently in building and implementing machine learning methods and then follow with examples of attacks that an adversary can exploit without having direct access to a trader's codebase.  The examples are our own, and are purely illustrative. 

\begin{table}[!hp]
\caption{Examples of Adversarial Attacks in Finance}\label{attackstable}
\begin{longtable}{|p{25mm}|p{30mm}|p{70mm}|}
\hline \thead{Failures} & \thead{Description} & \thead{Example} \\
\hline 
Reward Hacking 
& When training, the stated reward differs from a true reward. 
& A learning agent was trained to create a perfect hedge, however transaction costs were poorly modelled, leading to poor performance. \\
\hline 
Side Effects 
& A reinforcement learning system disrupts the environment by advancing its goal. 
& A model has learned an order execution strategy for an illiquid asset, but by executing this strategy, changes the dynamics of the order book significantly, leading to increased risk.\\
\hline 
Distributional Shifts 
& The system is trained on one environment, but unable to adapt to changes. & A pricing model was trained on data during normal times, and is unable to react to the higher correlations between assets during crises. \\
\hline 
Natural Adversarial Examples. 
& Even without being attacked the system fails from natural errors.
& A pricing model was trained individually for each strike and maturity, resulting in arbitrageable prices being offered in the market.  \\
\hline Common Corruptions
& The system is not able to deal with common corruptions. 
& A pricing model failed due to a halt on trading being placed on a closely related underlying instrument. \\
\hline Incomplete Testing 
& The system is not tested on the right environment nor over multiple periods. 
& A pricing model is tested only on one exchange, but is deployed in multiple locations with differing market behaviours. \\
\hline Poisoning attack 
& Contaminate training phase. 
& Contaminated data is introduced into a pricing model, for example when using sentiment analysis based on social media. \\
\hline Model stealing 
& Recover the entire model.
& A proprietary model is trained and can be queried online by counterparties. By repeated queries it is possible that the inputs can be matched with the outputs, to reverse engineer the original model. \\
\hline Model inversion 
& Recover hidden features.
& A pricing model is trained using proprietary trading data on market impact. The fitted model is then made public, without the underlying data. By repeated queries, it may be possible to extract the training data used \citep{fredrikson2015model}. \\
\hline Reprogramming system 
& Repurpose system for other use. 
& An online pricing model is used to identify expected future market volatility. \\
\hline
\end{longtable}
\end{table}

\begin{table}[!ht]
\begin{longtable}{|p{25mm}|p{30mm}|p{70mm}|}
\hline Adversarial example in physical domain
& Fool a system by changing some interface component.
& An adversary determines that a pricing model has sensitivity to the volumes deep in the order book -- by posting to this part of the book, they influence the model's behaviour.\\
\hline Exploit software dependencies
& The use of traditional software exploits.
& The model relies on code dependencies, these dependencies are exploited by modifying the code to introduce nonsensical values, leading to a trading halt. (The 2016 NPM/left-pad debacle illustrates this external dependency risk, where a disgruntled developer deleted a tiny piece of code that `broke' the internet \citep{Collins2021Feb}.) \\
\hline
\end{longtable}
\end{table}

These intentional and unintentional attack examples are hypothetical, and relate to problems seen in other machine learning domains. Nonetheless, these examples have significant implications for financial model risk management. A substantial level of compounded risks could exist where multiple of these susceptibilities overlap. 

Although there is a need to test and benchmark the robustness and resilience of trading agents with private systems and historical data, these agents ultimately have to move to the real world, where a slight distributional shift could impair performance. In other areas of machine learning, in addition to internal testing, models can be subjected to public audits. However, in finance the competitive risks from revealing private models are significant, leading to a far lower level of transparency.

\subsubsection{Model drift}

A good model not only fits historical data well, but also captures changes in the environments in which it is deployed. The challenge of updating models exists in both handcrafted and machine learning models, and reflects the basic challenge that finance does not operate according to stable physical laws, but arises from the interactions of many agents.

The challenge of changing market behaviour can be significant: the overwhelming belief is that the value of a derivative and its underlying are kept in line due to no-arbitrage. However, during the 2007-08 financial crisis, these relationships were observed to break down, as arbitrage calculations did not account for counterparty creditworthiness. As a result, a theoretical arbitrage opportunity was observable in the market, but was not available in practice \citep{baba2009interpreting}.

Handcrafted models, typically, require updates of few parameters to capture the shift of the data distribution. For overparameterised models, this may not be the case, and a small change in the data may require a significant change in the model.  For example: fraud detection models lose their discriminatory power against maliciously evolving strategies,  hedging strategies have to evolve as market conditions are changing.  

Off-line machine learning suffers from a lack of robustness to distribution shifts, and hence a lack of on-line monitoring can significantly impair its performance \citep{sugiyama2012machine}. This has become particularly clear in recent years in other applications of machine learning. For example, in the airline industry it was quickly realised that the standard machine learning pricing models that study flight patterns, fuel costs, and user behaviour became useless during the covid-19 pandemic, with data scientists choosing to fall back on traditional macroeconomic modelling \citep{McCartney2020Aug}.

On the other hand, online learning approaches have the promise of being able to dynamically and naturally adapt to new situations \citep{zeng2019online, soleymani2020financial}. This comes with significant issues, however, as these methods require training at a meta-level: the rate at which they adapt to new information needs to be tuned and adjusted, with rapid adjustment speeds typically associated with increased volatility in performance.

\subsection{Model solutions}
Any given model provides only a crude approximation to reality; the risk of using an inadequate model is often hard to detect and quantify.  While modern data science techniques are opening the door to more data-driven model selection mechanisms, this comes with new risks, as described previously.  In this section, we argue that by combining old and new approaches, it is possible to regain control over newly emerging risks (e.g. lack of interpretability) while improving over classical models currently favoured by industry. We base our presentation on a few hybrid modelling approaches which have recently emerged in the research literature. 

A natural idea is to incorporate prior knowledge/modelling into deep learning. This can be achieved through incorporating modelling constraints during the training. However, as the number of constraints increases, and hence the search space of possible network parameters decreases, stochastic gradient descent algorithms struggle to find good solutions, so bespoke machine learning methods need to be developed.

\subsubsection{Machine learning as a numerical tool}

As mentioned above, using machine learning as a numerical tool introduces only modest model risks, while potentially providing significant speed and accuracy benefits. In \citet{vidales2018unbiased,sabate2020solving}, the authors developed deep learning algorithms for solving parametric families of (path-dependent) partial differential equations (P)PDEs that arise in pricing and hedging.  The key idea in these works is to use a probabilistic representation of the (P)PDE, and learn both the solution and its gradient simultaneously. An advantage of this approach is that the gradient of the solution to the (P)PDE provides access to the hedging strategies. While this method is of interest in its own right, it can also be used as a control variate for unbiased Monte Carlo pricing.  In other words, by combining deep learning with standard Monte Carlo pricing, one can remove the bias due to approximation with neural nets and easily compute confidence intervals (which are, in general, hard to obtain for large networks).  This approach has been tested on several models and (path dependent) payoffs.  We stress that while the literature on deep learning for PDEs is growing rapidly, for finance applications it is critical to approximate parametric \emph{families} of PDEs, where parameters correspond to the possible values of calibrated coefficients of the model. A similar observation has been made in \citep{horvath2020deep}.

Another interesting approach, that combines ideas emerging from ML and classical modelling has been put forward in \citep{lyons2019nonparametric}. The key idea here is to lift both modelling and pricing into the signature space.  Intuitively, signatures provide efficient basis functions for representing functionals defined on the path space (e.g exotic derivatives or non-Markovian models)  and play a similar role to polynomials on Euclidean space. In particular, the signature expansion of a path represents the values of integrals against that path, and so can capture the effect of dynamic trading and hedging. The classical idea of replicating an option via trading in the market then reduces to regressing the option payoff on the signature of the underlying and other vanilla securities.

It has been shown that one can effectively represent many exotic derivatives using this signature expansion, and consequently obtain the prices of derivatives in terms of the expectation (under the pricing measure) of the  signature expansion terms. Consequently, one only needs to  calibrate expected signatures to market data, which in some settings can be done efficiently.  The advantage of using signatures when compared with recursive neural networks is that the computational cost does not increase with the number of time points in a time-series.

The idea of model selection using signatures has been proposed in \citep{arribas2020sig}. Here, one still works with the familiar SDEs type model but aims to learn (possibly non-Markovian) coefficients from data.

\subsubsection{Expert knowledge}

A viable approach to controlling the risk of non-transparent model specifications is to develop algorithms and training methods that embed expert knowledge into the architecture or training stage of machine learning. A handful of papers have attempted to embed financial domain knowledge into their models. These methods can offer regularisation, efficiency, consistency, and stability benefits. 

Drawing from the review by \citet{ruf2020neural}, methods that adjust the \emph{architectural} design of neural networks include models that incorporate a homogeneity hint by training a neural network in two parts, the first part controls for moneyness, and the other for time-to-maturity \citep{garcia1998option}. Other methods restrict the shape of outputs \citep{dugas2001incorporating} or enforce no-arbitrage conditions such as the convexity of a neural network pricing function and monotonicity \citep{zheng2019gated}. 

Approaches that impart expert knowledge at the \emph{training} stage include data augmentation, which involves the generation of synthetic data to help with neural network training \citep{yang2017gated}, adjustments in the penalty terms of the loss function to promote no-arbitrage \citep{itkin2019deep, ackerer2019deep}, as well as the development of bespoke training algorithms for neural networks for options hedging, including the use of the extended Kalman filter, sequential Monte Carlo, and evolutionary algorithms \citep{niranjan1996sequential,de2000hierarchical,palmer2019evolutionary}.

\subsubsection{Benchmarks}

A safe and efficient transition toward using machine learning in finance is only possible when models and methods are well understood and tested on reliable data sets. In other areas of machine learning, standard benchmarks and data sets are a common way to proof-test new methodologies. For example, recent advances in computer vision or reinforcement learning were significantly accelerated due to the emergence of challenging benchmarks, such as ImageNET \citep{deng2009imagenet} or ALE \citep{bellemare2013arcade}. These benchmarks have enabled open, systematic cross-validation of various AI solutions.

In machine learning, the term `benchmarking' has been used to refer to the evaluation and comparison of machine learning models, particularly regarding their ability to learn patterns from benchmark datasets \citep{olson2017pmlb}. This process can be thought of as a check to validate the improvement of a new method, but also more broadly to identify the respective advantages and disadvantages of each method. Comparisons can be made across a wide range of metrics, for example accuracy in detecting signals, interpretability, and computational complexity. 

Currently, in finance, various algorithms and machine learning methods are tested on disparate data sets, which are often only accessible to a small community or at high cost. A consequence of this is that very little comparison of methods is done, and we have little understanding of the appropriateness or optimality of these methods. In addition, evaluating new AI  techniques on real-world applications often requires expert domain knowledge and consideration of scalability and the cost of development. 

A key difficulty, in financial applications, is that a more open approach to benchmarking will often involve revealing details of each participant's methodologies. While this is reasonable within the academic community, within industry it is clear that confidentiality is needed, both regarding algorithms and, in some cases, their performance. For this reason, it is important to build our understanding of which problems can be discussed and benchmarked in a public way, and which related data science problems provide insight for those cases where confidentiality is needed.

The typical datasets which the benchmarking literature has well studied come from real-world data and simulated data with known underlying patterns. As alluded to before, in finance there are relatively few datasets that have been made publicly available, and often these contain only a small sample of the data that would be needed in practice. There is therefore a clear opportunity for Finance to benefit from synthetic data generators. Synthetic data has been used in other fields\footnote{ For example, the Open Graph Benchmark, released in 2020, has become a popular repository of challenging and realistic benchmark datasets to help facilitate scalable, robust, and reproducible graph machine learning research \citep{hu2020open}.} but has not yet flourished in the financial literature. 

Benchmarking has its own problems, many of which are not new to machine learning. There has been an increasing concern that published research findings are misleading due to the number of studies addressing the same question and datasets \citep{ioannidis2005most}. Benchmarking has a similar problem, in that a lot of models are prodding the same unchanging datasets leading to a lack of generalisation. Studies reveal that the accuracy of state-of-the-art deep learning models can drop from 4\%-10\% when moving to a new test-set, highlighting the risk of overfitting \citep{recht2018cifar}. For this reason, the regular evaluation and updating of benchmarks remains important for future development.

\subsubsection{Adversarial defenses}
In order to be reliably implemented, algorithms must be robust with respect to a variety of objectives (e.g. safety, accuracy).  Summarizing the range of adversarial challenges outlined above, we see that machine learning pipelines should come with robustness guarantees against: (i) shifts in data distribution (distributional robustness), (ii) intentional input manipulations (adversarial robustness) and (iii) intentional feature manipulation to ‘game’ the system (strategic robustness).

Recent work \citep{huang2017safety} has begun to address these issues for neural network based models. Drawing on adversarial machine learning and distributionally robust optimisation \citep{rahimian2019distributionally, cohen2019certified,wicker2020probabilistic}, it is possible to certifiably train models to provably ensure robustness, by providing guaranteed bounds on the probability of the model output (decision) satisfying a combination of objectives. 

Data driven models cannot automatically guarantee model robustness \citep{kwiatkowska2019safety}. An adversarial defense is anything that decreases the efficacy of adversarial attacks. There are a range of techniques that can be used to provide adversarial defenses; they can generally be classified into adversarial training methods, randomisation-based schemes, denoising methods, and provable defenses.

\begin{itemize}
    \item Adversarial learning techniques simply train a neural network using adversarial samples. It is one of the most effective defenses against attacks as revealed in benchmark studies \citep{madry2017towards}. These can be thought of as a preprocessing technique.  
    \item Randomisation schemes can also protect against perturbation in inputs. These generally involve some transformation, such as random resizing, or can also be achieved by adding a noise layer to the neural network \citep{liu2018towards}.
    \item Denoising inputs in the prediction phase can help to rectify or remove adversarial perturbations. This denoising can be done with generative adversarial networks or autoencoders and can be thought of as a postprocessing technique \citep{xie2019feature}.
    \item   Provable defenses are unlike the above approaches, in that they are theoretically proven, rather than purely being experimentally validated. These methods can certify a level of robustness before the prediction stage \citep{balunovic2019adversarial}.
\end{itemize}

The defenses listed here can only verify and protect a system against a limited number of attacks. Security vulnerability attacks will have to be dealt with using domain expertise, rather than relying on  generalist defense mechanisms. Adversarial defenses will not protect against a badly developed model, and appropriate fail-safe mechanisms and human oversight remain a critical part of implementation.

\subsubsection{Explainability}

Explainability allows for human oversight of machine learning to be carried out effectively, ensuring that model risk is understood and controlled. Understanding the causes behind performance is a common part of risk management -- for example, the `Profit and Loss attribution test', which forms part of the Fundamental Review of the Trading Book \citep{mar2021}, requires a bank's hypothetical profits using front-office pricing models to be explained against their back-office risk models and factors, as part of the validation of those risk models.

The understanding of models and their risks is a significant challenge in finance. The 2007-2008 financial crises demonstrated that copulas, especially those proposed by \citet{li2000default}, were underpowered to model the risks of CDOs, but yet still were too large and complex to be understood and critiqued by users. In contrast, machine learning models are overpowered, have shiny user-interfaces, but are even more obscure. Machine learning has been promoted in well-cited papers as a method for systemic risk analysis, without only limited discussion  of the risks of using machine learning and its lack of interpretability \citep{kou2019machine, aziz2019machine}. 

 Neural networks are not inherently explainable, as input features become entangled and compressed into a single value via repeated non-linear transformations of a weighted sums \citep{xie2020explainable}. Explainability can be improved by prima facie selecting a more interpretable `white box' model, that is, adopting models which intrinsically are easier to query and understand. Neural network models can be designed to be more interpretable through joint training \citep{hendricks2016generating,iyer2018transparency}  or including attention mechanisms \citep{bahdanau2014neural,devlin2018bert,anderson2018bottom}. 

Although these solutions apply for neural networks in general, they do not necessarily apply in a reinforcement learning framework. In this setting, rule-based \citep{verma2018programmatically, hein2017particle}, or hierarchical \citep{shu2017hierarchical} methods are available. The purpose of rule-based methods is to present the policies in high level human-readable language, e.g. IF-THEN sequences. Hierarchical methods  divide policies into simpler sub-tasks, each of which are separately more interpretable than a flat policy, and are therefore useful to explain individual decisions, i.e. they provide `local' interpretability.

The above interpretable models generally forgo some performance for enhanced comprehensibility. As a result, as performance is often the primary concern, explainability techniques which can be applied to a black-box model need to be identified. These techniques can be grouped under the name `post-hoc' explainability. 

The types of post-hoc explanation methods are broad and include perturbation analysis, gradient analysis, example based explanations, and surrogate-modelling for local and global explanations \citep{adadi2018peeking}. Different applications and tasks require a different balance between explainability and performance. 

`Deep' reinforcement learning is based on neural network models, and adds an additional layer of incomprehensibility to the modelling process \citep{mnih2013playing}. Reinforcement learning models are complex, but it is often possible to use interpretable surrogate models as a means of simplifying and representing their actions; this is often easier than developing inherently interpretable models \citep{puiutta2020explainable}. A range of surrogates are available for this purpose, and include genetic programming techniques \citep{hein2018interpretable}, causal DAGs \citep{madumal2019explainable} and the use of tree-based models to approximate predictions \citep{coppens2019distilling}. However, when using surrogate models for explainability, it is wise to keep the underlying model as simple as possible, in order to make it easier for a surrogate model to reproduce its outputs. 

\subsubsection{Monitoring and control}

Models not only have to be validated on historical data, i.e. benchmarked, they also have to be monitored and controlled when running 'live'. In machine learning, this is related to `concept drift', which refers to data distributions changing over time, leading to faulty predictions \cite{vzliobaite2016overview}. The hope is that, with online learning incorporated in the approach, models can self-diagnose and self-correct when this occurs, but this is not always the case. Continuous recalibration may not be possible in all settings, due to regulatory requirements and the cost of recalibration \citep{cohen2019switching}. A good survey of concept and data drift and how to deal with it can be found in \citet{gama2014survey}. The importance of monitoring, recalibrating and updating systems, and ensuring sufficient human control, is a key part of the implementation of most automated systems in practice. 

In a financial setting, we might also want to base the criteria for drift on the execution of other methods (for example handcrafted strategies) that are run in parallel as 'controls` for performance. This allows one to study those occasions in which the performance of controls differed significantly from the model, highlighting points of concern. 

Machine learning models need more extensive monitoring procedures than handcrafted approaches due to the various risks they come with. Nevertheless, the promise of improved performance, the flexibility of modelling, and the speed advantages associated with embracing these new technologies means that there is no doubt about their broad incorporation into many parts of the finance industry.

\bibliographystyle{apalike}
\bibliography{mybibliography}

\end{document}